\documentstyle[aps,prb,epsf,multicol,amsmath,graphics,psfrag]{revtex}

\begin{document}
\draft

\title{Realistic description of electron-energy loss spectroscopy for 
  one-Dimensional Sr$_2$CuO$_3$
}

\author{
  A.~H\"{u}bsch,$^{a}$ J.~Richter,$^{a}$ C.~Waidacher,$^{a}$ 
  K.~W.~Becker,$^{a}$ and W. von der Linden$^{b}$
}
\address{
  (a) Institut f\"{u}r Theoretische Physik,
  Technische Universit\"{a}t Dresden, D-01062 Dresden, Germany \\
  (b) Institut f\"{u}r Theoretische Physik, Technische Universit\"{a}t Graz, 
  Petersgasse 16, A-8010 Graz, Austria
}

\maketitle

\begin{abstract}
  We investigate the electron-energy loss spectrum of one-dimensional undoped 
  CuO$_{3}$ chains within an extended multi-band Hubbard model and an extended 
  one-band Hubbard model, using the standard Lanczos algorithm. Short-range 
  intersite Coulomb interactions are explicitly included in these models, and 
  long-range interactions are treated in random-phase approximation. The 
  results for the multi-band model with standard parameter values agree very 
  well with experimental spectra of Sr$_{2}$CuO$_{3}$. In particular, the 
  width of the main structure is correctly reproduced for all values of 
  momentum transfer. We find no evidence for enhanced intersite interactions 
  in Sr$_{2}$CuO$_{3}$.
\end{abstract}


\widetext
\begin{multicols}{2}
\narrowtext


 
One-dimensional systems are easy to conceive in theory but hard to find in
nature, and their experimental realization is restricted to few materials.
These include mesoscopic systems like single-wall 
nanotubes\cite{Ijima,Bockrath} or chains of metal 
atoms,\cite{Brune,Yanson,Segovia} and macroscopic systems with a strong 
anisotropy in one spatial direction. Among the latter, Sr$_{2}$CuO$_{3}$ has 
been in the focus of recent research. It contains separated chains of 
corner-sharing CuO$_{4}$ plaquettes, and is related to 
high-temperature superconducting compounds of higher dimensionality. 
Generally, the electronic properties of Sr$_{2}$CuO$_{3}$ are dominated by 
strong correlations of the valence holes. The low-dimensional character of 
magnetic excitations in this material manifests itself in magnetic 
susceptibility measurements that have been successfully described in terms of 
a one-dimensional spin-$\frac{1}{2}$ Heisenberg 
antiferromagnet.\cite{Ami,Motoyama,Kojima}



\begin{figure}[b]
  \begin{center}
    \scalebox{0.35}{
      \psfrag{q(A)}[1][0]{\Large $\;\;\;$ q(\AA$^{-1}$)}
      \psfrag{Energy loss (eV)}[1][0]{\Huge Energy loss (eV)}
      \psfrag{Loss function (arb. units)}[1][0]{
        \Huge Loss function (arb. units)
      }
      \psfrag{0.1}[1][0]{\LARGE $0.1$}
      \psfrag{0.2}[1][0]{\LARGE $0.2$}
      \psfrag{0.3}[1][0]{\LARGE $0.3$}
      \psfrag{0.4}[1][0]{\LARGE $0.4$}
      \psfrag{0.5}[1][0]{\LARGE $0.5$}
      \psfrag{0.6}[1][0]{\LARGE $0.6$}
      \psfrag{0.7}[1][0]{\LARGE $0.7$}
      \psfrag{0.8}[1][0]{\LARGE $0.8$}
      \psfrag{1}[1][0]{\LARGE $1$}
      \psfrag{2}[1][0]{\LARGE $2$}
      \psfrag{3}[1][0]{\LARGE $3$}
      \psfrag{4}[1][0]{\LARGE $4$}
      \psfrag{5}[1][0]{\LARGE $5$}
      \includegraphics{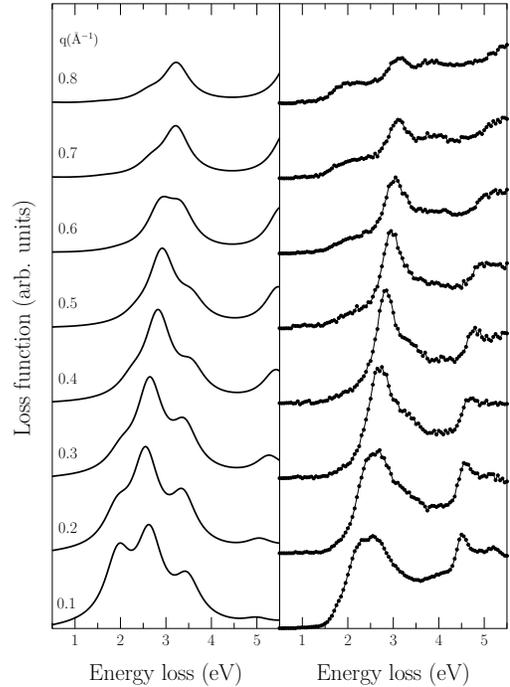}
    }
  \end{center}
  \caption{
    Comparison of experimental data for Sr$_2$CuO$_3$ (right panel), taken 
    from Ref.~\ref{Neudert}, and the results of the exact 
    diagonalization (left panel). The parameter set is $U_{d}=8.8\mbox{eV}$, 
    $\Delta=3.0\mbox{eV}$, $V_{pd}=1.2\mbox{eV}$, $V_{dd}=0\mbox{eV}$, 
    $t_{pd}=1.3\mbox{eV}$, and $t_{pp}=0.65\mbox{eV}$. The theoretical line 
    spectra have been convoluted with a Gaussian function of width 
    $0.35\mbox{eV}$.
  }
  \label{exp}
\end{figure}

Charge excitations in Sr$_{2}$CuO$_{3}$ have recently been 
investigated\cite{Neudert} by means of electron energy-loss spectroscopy 
(EELS). The experimental spectra are shown in the right panel of 
Fig.~\ref{exp}. In the following we will only discuss the spectral region 
below $4$eV energy loss. Excitations at higher energies probably involve Sr 
orbitals, which are not included in models for the Cu-O 
structure.\cite{Neudert} In the low energy region the experimental data show a 
broad dominant low-energy structure at 2.4 eV for momentum transfer 
${\bf q}=0.1$\AA $^{-1}$, which shifts to $3.2$eV for ${\bf q}=0.8$\AA$^{-1}$ 
[see Fig.~\ref{exp}]. The behavior of this structure as a function of 
momentum transfer is rather unusual: with increasing momentum transfer up to 
$0.4$\AA$^{-1}$ the width of the structure decreases but increases again for 
${\bf q}>0.4$\AA$^{-1}$. These spectra have been interpreted using an extended 
one-band Hubbard model,\cite{Neudert} an effective two-band Hubbard 
model,\cite{Penc} and a multi-band Hubbard model.\cite{Richter} Although the 
main difference between these models is just the elimination of the oxygen 
degrees of freedom, the results have been discussed controversially. In the 
one-band model the behavior of the low-energy feature has been interpreted as 
the transfer of spectral weight from a continuum of excitations to an exciton, 
formed due to a strongly enhanced intersite Coulomb repulsion 
$V$.\cite{Neudert,Stephan} Excitonic features have also been discussed in the 
strong coupling limit of an effective two-band Hubbard model.\cite{Penc} 
However, the used coupling strengths are not experimentally relevant. 
Therefore, no direct comparison to experimental data was possible in 
Ref.~\onlinecite{Penc}. In contrast, in the multi-band model the dispersion of 
the low-energy feature has been explained in terms of a shift from a rather 
delocalized Zhang-Rice singlet-like excitation to a more localized 
one.\cite{Richter} However, no intersite Coulomb repulsion was included in the 
Hamiltonian.

Up to now, all theoretical approaches failed to correctly describe the 
discussed decrease and increase in spectral width of the low-energy feature as 
a function of increasing momentum transfer: For small momentum transfer, the 
one-band model overestimates the experimentally observed width by a factor of 
about two.\cite{Neudert} In addition, the broadening for large momentum 
transfer is to small. The analytical approach to the multi-band model, on the 
other hand, underestimates the broadening due to the neglect of far reaching 
excitations that are important at small momentum transfer.\cite{Richter} 

In this paper, we show that the multi-band model provides a realistic 
description of the EELS spectrum for Sr$_{2}$CuO$_{3}$, and we observe the  
correct spectral form for all values of momentum transfer. Furthermore, it is 
found that the main effect of the intersite Coulomb repulsion is to lead to an 
energy shift of the EELS spectra. Finally, we discuss the relations of our 
results to the loss function of the one-band model.



We investigate the dielectric response of a one-dimensional extended 
multi-band Hubbard Hamiltonian at half-filling, i.e.\ a chain of 
corner-sharing CuO$_{4}$ plaquettes with one hole per Cu site. In the hole 
picture this Hamiltonian reads\cite{Emery,Emery1}
\begin{eqnarray}
H &=&\Delta\sum_{j\sigma}n^p_{j\sigma}
+U_{d} \sum_{i}n^d_{i\uparrow}n^d_{i\downarrow}\nonumber\\
&&+V_{pd} \sum_{<ij>}n^p_{j}n^d_{i}
+V_{dd} \sum_{<ii^\prime>}n^d_{i}n^d_{i^\prime}\nonumber\\
&&+t_{pd} \sum_{<ij>\sigma}\phi^{ij}_{pd}
(p^\dagger_{j\sigma}d_{i\sigma}+{\rm H.c.})\nonumber\\
&&+ t_{pp} \sum_{<jj^\prime>\sigma}\phi^{jj^\prime}_{pp}
p^\dagger_{j\sigma}p_{j^\prime\sigma}~\mbox{.}\label{1}
\end{eqnarray}
The operators $d^\dagger_{i\sigma}$ ($p^\dagger_{j\sigma}$) create a hole with 
spin $\sigma$ in the $i$-th Cu $3d$ orbital ($j$-th O $2p$ orbital), and 
$n^d_{i\sigma}$ ($n^p_{j\sigma}$) are the corresponding number operators, with
$n^d_{i}=n^d_{i\uparrow}+n^d_{i\downarrow}$. The first four terms in 
Eq.~(\ref{1}) are the atomic part of the Hamiltonian, with the charge-transfer 
energy $\Delta$, the Cu on-site Coulomb repulsion $U_{d}$, the Cu-O intersite 
repulsion $V_{pd}$, and the Cu-Cu intersite repulsion $V_{dd}$. The last two 
terms in Eq.~(\ref{1}) describe the hybridization of Cu $3d$ and O $2p$ 
orbitals (hopping strength $t_{pd}$) and of O $2p$ orbitals (hopping amplitude 
$t_{pp}$).  $\phi^{ij}_{pd}$ and $\phi^{jj^\prime}_{pp}$ give the correct sign 
for the hopping processes, and $\langle ij \rangle$ denotes the summation over 
nearest neighbor pairs. Hamiltonian \eqref{1} takes account of both in-chain 
and out of chain oxygen sites. Notice that no perturbative approximations are 
made, so that parameter values can be chosen in the experimentally relevant 
range.

The dynamical density-density correlation function is directly proportional to 
the loss function in EELS experiments.\cite{Schnatterly} By including the 
long-range Coulomb interaction in the model within a random-phase 
approximation\cite{Pines} (RPA) one finds for the loss function
\begin{eqnarray}
  L(\omega,{\bf q}) & = & 
    {\rm Im}\left[
      \frac{-1}{1+v_{\bf q}\chi_{\rho}^{0}(\omega,{\bf q})}
    \right],\label{lost}
\end{eqnarray}
where 
\begin{eqnarray}
  \chi_{\rho}^{0}(\omega,{\bf q}) & = &
    \frac{i}{\hbar}\int_{0}^{\infty}dt\,e^{i\omega t}\langle 
      0|[ \rho_{\bf q}(t), \rho_{-{\bf q}}]|0
    \rangle
\end{eqnarray}
is the response function at zero temperature of the short-range interaction 
model \eqref{1}. $\chi_{\rho}^{0}$ depends on the energy loss $\omega$ and 
momentum transfer ${\bf q}$. $|0\rangle$ is the ground state, $\rho_{\bf q}$ 
denotes the Fourier transform of $n_{i}$, and 
$v_{\bf q}=e^{2}N/(\epsilon_{0}\epsilon_{\rm r}v{\bf q}^{2})$ is the 
long-range Coulomb interaction with unit cell volume $v$. $N$ is the number of 
electrons per unit cell, and $\epsilon_{o}$ is the permittivity. The real part 
$\epsilon_{\rm r}$ of the dielectric function can be obtained from the 
experiment. In the case of Sr$_{2}$CuO$_{3}$ it was found to be  
$\epsilon_{\rm r}=8$.\cite{Neudert} In the following we evaluate 
Eq.~\eqref{lost} using the standard Lanczos algorithm\cite{Lin} which is 
limited to small clusters. The theoretical line spectra are convoluted with a 
Gaussian function of width $0.35$eV, to allow a comparison with the experiment.

\begin{figure}
  \begin{center}
    \scalebox{0.55}{
      \psfrag{q(A )}[1][0]{\large q(\AA$^{-1}$)}
      \psfrag{Energy loss (eV)}[1][0]{\Large Energy loss (eV)}
      \psfrag{Loss function (arb. units)}[1][0]{
        \Large Loss function (arb. units)
      }
      \psfrag{0.1}[1][0]{\large $0.1$}
      \psfrag{0.3}[1][0]{\large $0.3$}
      \psfrag{0.5}[1][0]{\large $0.5$}
      \psfrag{0.7}[1][0]{\large $0.7$}
      \psfrag{0}[1][0]{\large $0$}
      \psfrag{1}[1][0]{\large $1$}
      \psfrag{2}[1][0]{\large $2$}
      \psfrag{3}[1][0]{\large $3$}
      \psfrag{4}[1][0]{\large $4$}
      \psfrag{5}[1][0]{\large $5$}
      \includegraphics{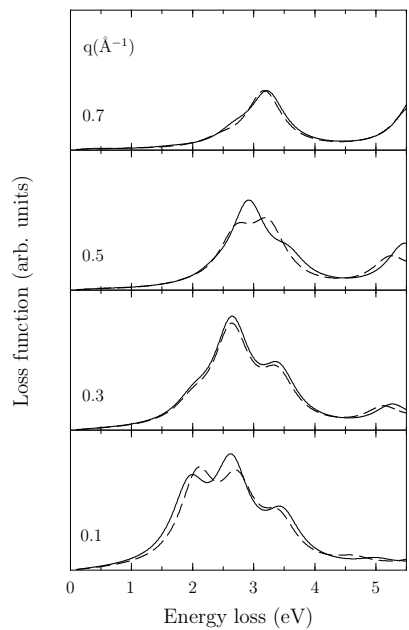}
    }
\end{center}
  \caption{
    Finite-size effects in the loss function of model \eqref{1} for clusters 
    with six plaquettes (full lines) and five plaquettes (dashed line), with 
    open boundary conditions. Parameters as for Fig.~\ref{exp}.
  }
  \label{size}
\end{figure}

First we check if our results are sufficiently converged with respect to 
system size. In Fig.~\ref{size} we compare the loss function of clusters with 
five (dashed lines) and six plaquettes (full lines). In both cases open 
boundary conditions are chosen. One has to make sure that holes on 
the edges of the cluster are still embedded in the local Coulomb potential 
that results from a state with occupied Cu sites. For this purpose, O (Cu) 
sites on the edge of the cluster are assigned an additional on-site energy 
due to $V_{pd}$ ($V_{dd}$). As can be seen from Fig.~\ref{size}, there are 
only small finite-size effects. Thus we conclude that the cluster with six 
plaquettes is large enough to obtain reliable results.

In Fig.~\ref{exp} the calculated loss function is compared to the experimental 
spectra from Ref.~\onlinecite{Neudert}. The parameters in the model 
Hamiltonian are chosen as follows: $U_{d}=8.8$eV, $V_{pd}=1.2$eV, 
$V_{dd}=0$eV, $t_{pd}=1.3$eV and $t_{pp}=0.65$eV are kept constant at typical 
values.\cite{McMahan,Hybertsen,Grant} The value of $\Delta=3.0$eV has been 
adjusted to obtain correct peak positions. This means that we use only one 
free parameter. As compared to the standard value $3.5$eV for 
$\Delta$,\cite{McMahan,Hybertsen,Grant} the smaller $\Delta$ is in agreement 
with theoretical analysis of x-ray photoemission spectra for 
Sr$_{2}$CuO$_{3}$.\cite{Okada,Waidacher,Waidacher1} Notice that a small value 
of $\Delta$ means that the system is not in the strong coupling limit as was 
assumed in Ref.~\onlinecite{Penc}. The calculated loss function consists of a 
dominant structure at $2.5$eV for ${\bf q}=0.1$\AA$^{-1}$, which shifts to 
$3.2$eV for ${\bf q}=0.8$\AA$^{-1}$. Besides, a second excitation is observed 
at $5.5$eV. In agreement with the experimental observation, with increasing 
momentum transfer the main structure shifts to higher energies and first 
decreases in width. For ${\bf q}>0.4$\AA$^{-1}$ the spectral width increases 
again. The main structure results from excitations in which a hole leaves its 
original plaquette to form Zhang-Rice singlet-like states\cite{Zhang} with 
neighboring holes. With increasing momentum transfer {\bf q} the spectral 
weight shifts from extended to more localized excitations.\cite{Richter} 

\begin{figure}
  \begin{center}
    \scalebox{0.55}{
      \psfrag{Energy loss (eV)}[1][0]{\Large Energy loss (eV)}
      \psfrag{Loss function (arb. units)}[1][0]{
        \Large Loss function (arb. units)
      }
      \psfrag{(a1)}[1][0]{\large (a$_{1}$)}
      \psfrag{(b1)}[1][0]{\large (b$_{1}$)}
      \psfrag{(c1)}[1][0]{\large (c$_{1}$)}
      \psfrag{(a2)}[1][0]{\large (a$_{2}$)}
      \psfrag{(b2)}[1][0]{\large (b$_{2}$)}
      \psfrag{(c2)}[1][0]{\large (c$_{2}$)}
      \psfrag{0}[1][0]{\large $0$}
      \psfrag{1}[1][0]{\large $1$}
      \psfrag{2}[1][0]{\large $2$}
      \psfrag{3}[1][0]{\large $3$}
      \psfrag{4}[1][0]{\large $4$}
      \psfrag{5}[1][0]{\large $5$}
      \includegraphics{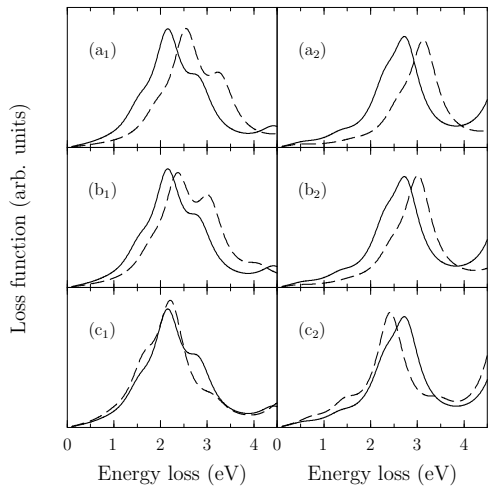}
    }
  \end{center}
  \caption{
    Influence of $V_{pd}$, $\Delta$, and $V_{dd}$ on the loss functions with 
    momentum transfer ${\bf q}=0.3$\AA$^{-1}$ (left side) and 
    ${\bf q}=0.7$\AA$^{-1}$ (right side); other parameters as for 
    Fig.\ref{exp}. In (a) $\Delta=3.0$eV and $V_{dd}$ are kept constant and 
    $V_{pd}$ is varied (full line: $V_{pd}=0$, dashed line: $V_{pd}=1.0$eV), 
    in (b) $\Delta$ is varied (full line: $\Delta=3.0$eV, dashed line: 
    $\Delta=4.0$eV) for $V_{pd}=V_{dd}=0$, and in (c) $\Delta=3.0$eV and 
    $V_{pd}=0$ are constant (full line: $V_{dd}=0$, dashed line: 
    $V_{dd}=0.5$eV). The loss functions with ${\bf q}=0.3$\AA$^{-1}$ and 
    ${\bf q}=0.7$\AA$^{-1}$ are scaled independently of each other.
  }
  \label{Vpd}
\end{figure}

Next we discuss the dependence of the calculated spectra on the different 
parameters in Hamiltonian \eqref{1}. In agreement with an analysis \cite{Okada}
of the optical conductivity for Sr$_{2}$CuO$_{3}$ it is found that the 
main influence of $\Delta$ and the intersite Coulomb repulsion $V_{pd}$ is a 
shift of the excitation energy of the main structure. This is shown in 
Fig.~\ref{Vpd} for $V_{pd}$ [see panels (a$_{1}$) and (a$_{2}$)] and $\Delta$ 
[see panels (b$_{1}$) and (b$_{2}$)]. Note that we observe nearly the same 
results for $\Delta=3.0$eV, $V_{pd}=1.0$eV [dashed lines in panels (a$_{1}$) 
and (a$_{2}$) of Fig.~\ref{Vpd}] and $\Delta=4.0$eV, $V_{pd}=0$ [dashed lines 
in panels (b$_{1}$) and (b$_{2}$) of Fig.~\ref{Vpd}]. Hence, only the sum of 
both parameters $\Delta+V_{pd}$ is relevant for the spectra. Therefore, it is 
possible to obtain good agreement between experimental and theoretical spectra 
with or without intersite Coulomb interaction $V_{pd}$. This implies that the 
mechanism of excitations is not driven by a strong intersite interaction 
$V_{pd}$.  Furthermore, this observation also explains why a fit of a 
multi-band model with $V_{pd}=0$ to the experimental data has led to a larger 
value of $\Delta$ in Ref.~\onlinecite{Richter}.

\begin{figure}
  \begin{center}
    \scalebox{0.55}{
      \psfrag{Energy loss (eV)}[1][0]{\Large Energy loss (eV)}
      \psfrag{Loss function (arb. units)}[1][0]
        {\Large Loss function (arb. units)}
      \psfrag{(a1)}[1][0]{\large (a$_{1}$)}
      \psfrag{(b1)}[1][0]{\large (b$_{1}$)}
      \psfrag{(a2)}[1][0]{\large (a$_{2}$)}
      \psfrag{(b2)}[1][0]{\large (b$_{2}$)}
      \psfrag{0}[1][0]{\large $0$}
      \psfrag{1}[1][0]{\large $1$}
      \psfrag{2}[1][0]{\large $2$}
      \psfrag{3}[1][0]{\large $3$}
      \psfrag{4}[1][0]{\large $4$}
      \psfrag{5}[1][0]{\large $5$}
      \includegraphics{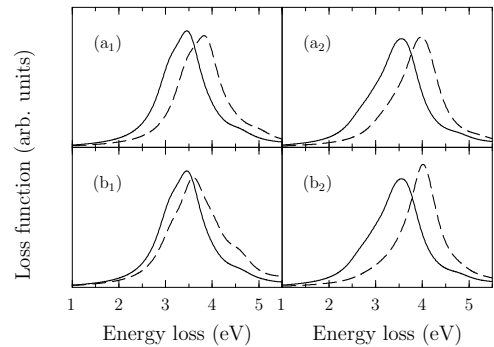}
    }
\end{center}
  \caption{
    Influence of the Coulomb interactions $U$ and $V$ of the extended one-band 
    Hubbard model on the loss function with momentum transfer 
    ${\bf q}=0.3$\AA$^{-1}$ (left side) and ${\bf q}=0.7$\AA$^{-1}$ (right 
    side); parameter values are chosen according to Ref.~\ref{Neudert}. The 
    hopping strength is $t=0.55$eV. In (a) $V=1.3$eV is kept constant (full 
    line: $U=4.2$eV, dashed line: $U=4.7$eV), and in (b) $V$ is varied (full 
    line: $V=1.3$eV, dashed line: $V=0.8$eV) for $U=4.2$eV. The theoretical 
    line spectra have been convoluted with a Gaussian function of width 
    $0.35$eV. The loss functions with ${\bf q}=0.3$\AA$^{-1}$ and 
   ${\bf q}=0.7$\AA$^{-1}$ are scaled independently of each other.
  }
  \label{hubb}
\end{figure}

In contrast to $\Delta$ and $V_{pd}$, the Cu-Cu intersite repulsion $V_{dd}$ 
does not shift the complete EELS spectra but rather transfers spectral weight 
to excitations with smaller energy loss [see panels (c$_{1}$) and (c$_{2}$) of 
Fig.~\ref{Vpd}]. A comparison of panels (c$_{1}$) and (c$_{2}$) shows that 
this effect is larger near the zone boundary at ${\bf q}=0.8$\AA$^{-1}$. 
Probably, this behavior can be connected with a formation of an exciton state 
as discussed in Ref.~\onlinecite{Neudert,Penc}. On the other hand, the choice 
$V_{dd}=0$ is a good approximation\cite{Dagotto} for the multi-band model 
\eqref{1} since the distance between neighboring Cu sites is relatively large. 
Therefore, a possible exciton formation seems not to be relevant for the 
interpretation of the experimental spectra if one uses the multi-band model.

Finally, we want to discuss the relation of our results for the multi-band 
model to the loss function \eqref{lost} of the one-dimensional extended 
Hubbard model
\begin{eqnarray}
  H &=& -t\sum_{\langle ij\rangle\sigma}(d_{i\sigma}^{\dagger}d_{j\sigma} + 
           {\rm H.c.}) + U\sum_{i}n_{i\uparrow}^{d}n_{i\downarrow}^{d} + 
           V\sum_{\langle ij\rangle}n_{i}^{d}n_{j}^{d}\nonumber\\
  && \label{ext_Hubb}
\end{eqnarray}
which only considers effective Cu $3d$ orbitals. In Eq.~\eqref{ext_Hubb}, $t$ 
is the hopping strength, $U$ denotes the on-site Coulomb repulsion, and $V$ is 
the intersite interaction. Note that the mapping of the multi-band model 
\eqref{1} onto the one-band model \eqref{ext_Hubb} is problematic for the 
model parameters used above because the system is not in the strong coupling 
limit. Therefore, we may compare the loss function of both models only 
qualitatively. In the following, we compute the loss function \eqref{lost} of 
the one-band model \eqref{ext_Hubb} using a cluster with twelve sites and 
periodic boundary conditions. If one reduces the multi-band model \eqref{1} to 
a one-band model the charge-transfer gap $\Delta$ is replaced by the Hubbard 
gap $U$ of the effective model.\cite{Zaanen} Therefore, in analogy to the 
influence of $\Delta$ in the multi-band model, increasing $U$ shifts the 
spectra to higher energy loss [see panels (a$_{1}$) and (a$_{2}$)of 
Fig.~\ref{hubb}]. In panels (b$_{1}$) and (b$_{2}$) of Fig.~\ref{hubb} the 
loss functions with ${\bf q}=0.3$\AA$^{-1}$ and ${\bf q}=0.7$\AA$^{-1}$ are 
shown for $V=1.3$eV (full lines) and $V=0.8$eV (dashed lines) where $U=4.2$eV 
and $t=0.55$eV are kept constant at the values from Ref.~\onlinecite{Neudert}. 
In analogy to the influence of the intersite Coulomb repulsion $V_{dd}$ in the 
multi-band model discussed above, a moderate increase in $V$ leads mainly to a 
transfer of spectral weight to excitations with smaller energy loss. However, 
a non-zero intersite Coulomb repulsion $V$ is needed to obtain spectra related 
to the experiment. This fact led to the conclusion that the spectral intensity 
at the zone boundary is due to an exciton formation.\cite{Neudert} On the 
other hand, the large differences between the interpretations of the 
experimental spectra for Sr$_{2}$CuO$_{3}$ using the one-band\cite{Neudert} 
and the multi-band model imply that oxygen degrees of freedom are important 
for a realistic description of charge excitations in the cuprates.



In conclusion, we have carried out an investigation of the EELS spectrum for 
the one-dimensional CuO$_{3}$ chain using an extended multi-band Hubbard model 
and an extended Hubbard model. Our results for the multi-band model show very 
good agreement with experimental data for Sr$_{2}$CuO$_{3}$. In contrast to 
former investigations, we can explain the width of the main structure for all 
values of momentum transfer. For the multi-band model, only a combination of 
intersite Coulomb interaction and charge-transfer energy is relevant for the 
loss function. Consequently, we find no evidence for enhanced intersite 
interactions in Sr$_{2}$CuO$_{3}$. The different explanations for the spectral 
intensity at the zone boundary using the one-band and the more realistic 
multi-band model shows that the oxygen degrees of freedom are important for 
the description of charge excitations.



We would like to acknowledge fruitful discussions with S. Atzkern, S.-L. 
Drechsler, J. Fink, M. S. Golden, R. E. Hetzel, R. Neudert, and H. Rosner. 
This work was supported by DFG through the research programs GK 85 and SFB 463.
The calculations were performed on the Origin 2000 at Technische 
Universit\"{a}t Dresden.



\end{multicols}

\begin{references}

\bibitem{Ijima}S. Ijima, Nature {\bf 345}, 56 (1991). 
\bibitem{Bockrath}M. Bockrath, D. H. Cobden, J. Lu, A. G. Rinzler, 
         R. E. Smalley, L. Balents, and P. L. McEuen, Nature {\bf 397}, 598 
        (1999).
\bibitem{Brune}H. Brune, M. Giovannini, K. Bromann, and K. Kern, Nature 
         {\bf 394}, 451 (1998).
\bibitem{Yanson} A. I. Yanson, G. Rubio Bollinger, H. E. van den Brom, 
         N. Agrat, and M. van Ruitenbeek, Nature {\bf 395}, 783 (1998).
\bibitem{Segovia} P. Segovia, D. Purdie, M. Hengsberger, and Y. Baer, Nature 
         {\bf 402}, 504 (1999).
\bibitem{Ami}T. Ami, M. K. Crawford, R. L. Harlow, Z. R. Wang, D. C. Johnston,          Q. Huang, and R. W. Erwin, Phys. Rev. B {\bf 51}, 5994 (1995).
\bibitem{Motoyama}N. Motoyama, H. Eisaki, and S. Uchida, Phys. Rev. Lett. 
         {\bf 76}, 3212 (1996).
\bibitem{Kojima}K. M. Kojima, Y. Fudamoto, M. Larkin, G. M. Luke. J. Merrin, 
         B. Nachumi, Y. J. Uemura, N. Motoyama, H. Eisaki, S. Uchida, 
         K. Yamada, Y. Endoh, S. Hosoya, B. J. Sternlieb, and G. Shirane, 
         Phys. Rev. Lett. {\bf 78}, 1787 (1997).
\bibitem{Neudert}R. Neudert, M. Knupfer, M. S. Golden, J. Fink, W. Stephan, 
         K. Penc, N. Motoyama, H. Eisaki, and S. Uchida, Phys. Rev. Lett. 
         {\bf 81}, 657 (1998).\label{Neudert}
\bibitem{Penc}K. Penc and W. Stephan, Phys. Rev. B {\bf 62}, 12707 (2000).
\bibitem{Richter}J. Richter, C. Waidacher, and K. W. Becker, Phys. Rev. B 
         {\bf 61}, 9871 (2000).
\bibitem{Stephan}W. Stephan and K. Penc, Phys. Rev. B {\bf 54}, R17269 (1996).
\bibitem{Emery}V. J. Emery, Phys. Rev. Lett. {\bf 58}, 2794 (1987).
\bibitem{Emery1}V. J. Emery and G. Reiter, Phys. Rev. B {\bf 38}, 4547 (1988).
\bibitem{Schnatterly}S. E. Schnatterly, Solid State Phys. {\bf 34}, 275 (1977).
\bibitem{Pines}D. Pines and D. Bohm, Phys. Rev. {\bf 85}, 338 (1952).
\bibitem{Lin}For example, see H. Q. Lin and J. E. Gubernatis, Computer in 
         Physics {\bf 7}, 400 (1993), and references therein.
\bibitem{McMahan}A. K. Mahan, R. M. Martin, and S. Satpathy, Phys. Rev. B  
        {\bf 38}, 6650 (1988).
\bibitem{Hybertsen}M. S. Hybertsen, M. Schl\"{u}ter, and N. E. Christiansen, 
         Phys. Rev. B {\bf 39}, 9028 (1989).
\bibitem{Grant}J. B. Grant and A. K. McMahan, Phys. Rev. B {\bf 46}, 8440 
        (1992).
\bibitem{Okada}K. Okada, A. Kotani, K. Maiti, and D. D. Sarma, J. Phys. Soc. 
        Jpn. {\bf 65}, 1844 (1996).
\bibitem{Waidacher}C. Waidacher, J. Richter, and K. W. Becker, Europhys. Lett. 
        {\bf 47}, 77 (1999).
\bibitem{Waidacher1}C. Waidacher, J. Richter, and K. W. Becker, Phys. Rev. B 
        {\bf 61}, 13473 (2000).
\bibitem{Zhang}F. C. Zhang and T. M. Rice, Phys. Rev. B {\bf 37}, 3759 (1988).
\bibitem{Okada}K. Okada and A. Kotani, J. Phys. Soc. Jpn. {\bf 66}, 341 (1997).
\bibitem{Dagotto}E. Dagotto, Rev. Mod. Phys. {\bf 66}, 763 (1994).
\bibitem{Zaanen}J. Zaanen, G. A. Sawatzky, and J. W. Allen, Phys. Rev. Lett. 
         {\bf 55}, 418 (1985).

\end{references}
\end{document}